# Effect of lattice volume and strain on the conductivity of BaCeY-oxide ceramic proton conductors

Qianli Chen[1,2,*], Artur Braun[1], Songhak Yoon[3], Nikolai Bagdassarov[4], Thomas Graule[1,5]

[1] *Laboratory for High Performance Ceramics,*
*Empa, Swiss Federal Laboratories for Materials Science and Technology*
*CH-8600 Dübendorf, Switzerland*

[2] *Department of Physics, ETH Zürich - Swiss Federal Institute of Technology,*
*CH-8057 Zürich, Switzerland*

[3] *Laboratory for Solid State Chemistry and Catalysis,*
*Empa, Swiss Federal Laboratories for Materials Science and Technology*
*CH-8600 Dübendorf, Switzerland*

[4] *Institute for Geosciences, J. W. Goethe Universität Frankfurt am Main,*
*D-60438 Frankfurt/Main, Germany*

[5] *Institute of Ceramic, Glass- and Construction Material,*
*TU Bergakademie Freiberg, D-09596 Freiberg, Germany*

[*] Corresponding author: Phone +41 44 823 4129, Fax: +41 44 823 4150,
E-mail: qianlichen1@gmail.com

ABSTRACT

In-situ electrochemical impedance spectroscopy was used to study the effect of lattice volume and strain on the proton conductivity of the yttrium-doped barium cerate proton conductor by applying the hydrostatic pressure up to 1.25 GPa. An increase from 0.62 eV to 0.73 eV in the activation energy of the bulk conductivity was found with increasing pressure during a unit cell volume change of 0.7%, confirming a previously suggested correlation between lattice volume and proton diffusivity in the crystal lattice. One strategy worth trying in the future development of the ceramic proton conductors could be to expand the lattice and potentially lower the activation energy under tensile strain.

## 1. Introduction

Ceramic proton conductors with ABO$_3$-structure are recognized as potential candidates for electrolytes in future intermediate temperature solid oxide fuel cells.[1, 2] Among the best studied materials, Ba(Ce,Y)O$_3$ with Y-substitution at the B-side exhibits outstanding conductivity[3, 4] as a result from the proton motion in the lattice. The proton transport kinetics in the crystal lattice of these materials is closely related to the oxygen vacancies created by the Y$^{3+}$ substituent. These vacancies can be filled with the oxygen from adsorbed water molecule[5] and the introduced protons are bound to the lattice oxygen. Thermal activation acts as the driving force for the protons to overcome an energy barrier of a few hundred eV and jump to another oxygen site[6-8] and thus constituting proton mobility. The relationship of structure parameters and proton transport properties had been studied by Scherban[9] and Ashok[10] for example. Samgin[11] interpreted the proton mobility based on a result of phonon-activated jumps, where the O-O distance plays an important role in proton diffusion. Azad et al.[12] found by neutron diffraction that two cubic phases coexist in BaZr$_{0.9}$Y$_{0.1}$O$_{3-\delta}$ (BZY10) – a well conductive phase with larger lattice parameter, and a poor conductive phase with smaller lattice parameter. Wakamura[13] plotted the activation energy ($E_a$) versus lattice volume ($V$) for numerous ion and proton conductors, and found a simple inverse correlation between them according to the dominated force $E_a = A_v / V^{2/3}$, where $A_v$ is a constant. In our earlier work we have found a clear linear relationship between the $E_a$ for bulk proton conductivity and the lattice parameter for BZY10 with nominal identical stoichiometry but obtained by different synthesis routes[14]. A conceptually straightforward method to proof the correlation between lattice volume and transport properties is to apply mechanical pressure during the proton conductivity measurement. Our recent work[15] on the conductivity of BZY10 showed that the $E_a$ scales with the applied pressure. In this study, yttrium-substituted BaCeO$_3$ under compressive strain was investigated by in-situ electrochemical impedance spectroscopy in order to elucidate the effect of lattice volume and strain on the proton conductivity.

## 2. Experimental details

The BaCe$_{0.8}$Y$_{0.2}$O$_{3-\delta}$ (BCY20) powder was prepared by solid-state reaction. Stoichiometric amounts of BaCO$_3$ (Sigma-Aldrich, 99%), CeO$_2$ (Aldrich, 99.9%) and Y$_2$O$_3$ (Stanford Materials, 99.9%) were ball-milled in acetone with ZrO$_2$ balls of 10 mm diameter in a planetary mill (200 rpm) for 1 hour and calcined at 1000 °C for 12 hours successively for 2 times. The resulting powder was ball-milled again in a planetary mill and then axially pressed at 50 MPa for 0.5 minutes into pellets and sintered at 1400 °C for 24 hours. For protonation, sintered pellets or powder was heated to 500 °C in humid N$_2$ flow for 24 hours, resulting in proton concentration equivalent to [OH]$^-$ = 3.8 mol%.

Powder X-ray diffraction (PXRD) pattern was obtained in a Bragg–Brentano geometry using a PANalytical X´Pert PRO θ-2θ scan system equipped with Ge (111) monochromator and X'Celerator linear detector. The incident X-rays had a wavelength of 1.540598 Å (Cu-Kα$_1$). The diffraction patterns were scanned from 20° to 120° (2θ) with an angular step interval of 0.017°. XRD patterns have been analyzed by the Rietveld refinement program, *Fullprof*[16] to determine the lattice parameters, atomic positions and thermal factors. Thompson-Cox-Hastings (TCH) pseudo-Voigt functions were chosen as profile function among all the profiles in the *Fullprof* program[17].

The ionic conductivity was measured by electrochemical impedance spectroscopy (EIS) (Solartron 1260 Phase-Gain-Analyzer and Solartron 1296 Dielectric Interface) in the frequency range of 0.1 Hz – 3 MHz. Measurements under ambient pressure were performed in a ProboStat (NorECs AS) cell for a BCY20 pellet of 17 mm in diameter and 20 mm in thickness with painted Pt-paste contacts under humidity conditions[14, 18, 19]. In-situ high pressure EIS were measured for sintered powder at pressures ranging from 0.5 – 1.25 GPa in a piston-cylinder apparatus[20] from ambient temperature up to 440 °C. The EIS spectra were simulated in Software *Zview* and *Novocontrol WinFit* by an equivalent circuit model composed of three parallel RC series of the bulk, grain boundary and electrode responses.[14]

In order to monitor the loss of water at high temperature, the protonated sample was studied by thermogravimetric analysis (TGA) with Netzsch STA 409 under synthetic air, with a heating rate of 5°C/min up to 900 °C.

## 3. Results and discussions

Fig. 1 shows the Powder-XRD pattern and Rietveld refinement analysis of BCY20. The Rietveld refinement confirms that the BCY20 has monoclinic crystal structure with *I 2/m* space group, which was reported to be stable structure up to 500°C.[21, 22] The refined profile parameters and full crystallographic details are given in Table 1.

Fig.2 presents an example of three Nyquist plots at 125 °C at the pressures of 0.5, 0.75 and 1.0 GPa in which the impedance varies with pressure. The impedance spectra show typically a low impedance semicircle representing the bulk contribution, as well as a much higher impedance semicircle of grain boundary contribution. The third and largest semicircle originates from the Pt-paste electrodes. The TGA measurement in the inset of Fig. 2 confirms that there is no water loss up to 600°C, which is above the maximum temperature for the in-situ EIS, ensuring that the proton concentration did not change during the measurement. We can safely assume that for the temperatures where we recorded EIS spectra, the dominant contribution to the ionic conductivity originates from protons, and not from oxygen or oxygen vacancies. The onset of these latter species comes at significantly higher temperatures.[23] The impedance spectra were fitted with the equivalent circuit model of capacitance and resistivities (shown as an inset in Fig.2), based on which the resistance R and, together with the geometrical dimensions (sample thickness L, electrode surface area A), the conductivity for grain boundaries and bulk were determined:

$$\sigma = \frac{L}{A}\frac{1}{R} \tag{1}$$

The temperature and pressure dependence of the bulk, $\sigma_{bulk}$ and grain boundary $\sigma_{gb}$, conductivities are shown in Arrhenius plot in Fig.3(a) and (b). The $\sigma_{gb}$ is lower than $\sigma_{bulk}$, which is typically observed in ceramic ion conductors. The conductivity measured from a pellet at ambient pressure is also plotted for comparison. The values are generally one order of magnitude higher than those measured from the compressed powder, because of the higher density of the sintered pellet. For ambient pressure the significant change in activation energy comes from the difference in the proton jump rotation mode at low temperature and the diffusion mode at higher temperature.[6] We notice that the $\sigma_{gb}$ decreases as a function of pressure.

Fig. 4 illustrates the relationship between pressure and the activation energy $E_a$. During the pressure change from 0.5 to 1.25 GPa, the $E_a$ increases from 0.62 eV to 0.73 eV, and from 0.56 to 0.89 eV, respectively for the bulk and grain boundary region. The effect of pressure is more significant for the grain boundary conductivity than for the bulk conductivity, supporting the general perception that improving the grain boundary conductivity can be an effective pathway to control the total conductivity.

The lattice volume for BCY20 under pressure can be calculated by the equation of state obtained from the pressure – lattice volume relationship (Birch-Murnaghan equation of state with B'=4) at ambient temperature of a similar stoichiometric sample[24]:

$$p = \frac{3}{2} \cdot B \cdot \left( x^{-\frac{7}{3}} - x^{-\frac{5}{3}} \right) \qquad (2)$$

with the volume ratio $x=V/V_o$ and the bulk modulus $B = 103$ GPa (this value is taken from [24] for $BaCe_{0.85}Y_{0.15}O_{2.925}$ *in lieu* for BCY20 for which no data were available).

Fig. 5(a) shows the effect of unit cell volume on the proton conductivity $E_a$ in the scale reproduced after Wakamura[13]. The unit cell volume changed to 336.72 Å$^3$ at the pressure of 1.25 GPa, calculated from Eq. (1). Both the bulk and the grain boundary proton conductivity can be greatly influenced by small change in the lattice volume (0.7%). The observed effect is actually more significant than described in Ref. 13. For comparison, $E_a$ data from compressed BZY10[15] are plotted as well. The $E_a$ for proton hopping decreases with the increasing unit cell volume, which is derived from the weakened long-range force caused by the volume expansion from the host lattice. The lattice volume effect may not necessarily be the main and the only reason for the change of $\sigma_{gb}$. The structure and morphology of grain may also inflect on the transport properties.

For device applications, it is interesting but technically challenging to expand the lattice volume instead of compressing. According to the variation of $E_a$ as a function of elastic strain $\varepsilon = (l-l_0)/l_0$ revealed in Fig. 5(b), materials under tensile strain could have potentially lower activation energies and thus improved proton conductivity. Thin film technology, such as epitaxial growth could thus help to realize this idea, and minimize the high resistive grain boundary as well. As a future work it is highly anticipated that the proton conducting phases could be optimized with tensile strain by suitable sample preparation methods.

## 4. Conclusions

We investigated the hydrostatic pressure introduced changes of the cell parameters dependence of the conductivity in the proton-conducting perovskite $BaCe_{0.8}Y_{0.2}O_{3-\delta}$. The steady increase of the proton conductivity activation energy upon pressurizing suggests that the proton conductivity depends on the space available in the lattice. In return, expanded crystal lattices favor enhanced conductivity with decreasing the activation energy for proton conductivity, suggesting that thin films under expansive tensile strain could have larger proton conductivity with desirable properties for applications, such as proton conductor electrolytes and separation membranes.

**Acknowledgement**

This work was financially supported by E.U. MIRG Grant No. CT-2006-042095 and Swiss National Science Foundation Grant No. 200021-124812. The authors are gratefully debited to Thierry Strässle (Paul Scherrer Institut) for discussions about the equation of state, and to Hans-Jürgen Schindler (Empa) for the TGA measurements.

**Table 1.** Refined crystal structural parameters for BCY20 powder.

Fig. 1. Powder XRD pattern and Rietveld refinement profile of protonated BCY20. The difference plots of observed and calculated diffraction profiles are shown below together with the allowed Bragg positions in short vertical markers.

Fig. 2. Nyquist plots at 125 °C under different pressures and one representative fit; inset: mass loss up to 1000 °C; the equivalent circuit used for the impedance spectra modeling.

Fig. 3. Arrhenius plots of pressure and temperature dependent electrical conductivity for protonated BCY20, 3D- (a) and 2D-plot (b).

Fig. 4. The activation energy of bulk and grain boundary conductivities.

Fig. 5. Variation of activation energy of bulk conductivity on the lattice parameter under unit cell volume, together with other superionic conductors (SIC) and $H^+$-ion conductors (HIC); inset: in the volume scale of BCY20 (a); and on the elastic strain parameter $\varepsilon$ (b). BZY10 data are from Ref. 15.

**Table 1. Qianli Chen** *et al.*

Table 2. Refined crystal structural parameters for BCY20 powder.

| Name | \ | \ | \ | \ |
|---|---|---|---|---|
| Name | $BaCe_{0.8}Y_{0.2}O_{3-\delta}$ | | | |
| Radiation source | Lab x-ray $CuK_{\alpha 1}$ | | | |
| Wavelength (Å) | 1.540598 | | | |
| Temperature (K) | 298 | | | |
| 2θ range (°) | 20 - 120 | | | |
| Space Group | *I2/m* | | | |
| *a* (Å) | 6.23467 | | | |
| *b* (Å) | 8.73790 | | | |
| *c* (Å) | 6.25545 | | | |
| *β*(deg) | 91.00897 | | | |
| *V* (Å³) | 340.731 | | | |
| atom | x | y | z | $B_{iso}$ (Å²) |
| Ba | 0.25260 | 0 | 0.74698 | 1.14760 |
| Ce/Y | 0.25 | 0.25 | 0.25 | 0.66208 |
| O1 | 0.20724 | 0 | 0.23800 | 0.98593 |
| O2 | 0 | 0.31612 | 0 | 1.78554 |
| O3 | 0.5 | 0.21834 | 0 | 0.48056 |
| $R_p$ | 7.93 | | | |
| $R_{wp}$ | 10.5 | | | |
| $\chi^2$ | 2.24 | | | |

**Figure 1**

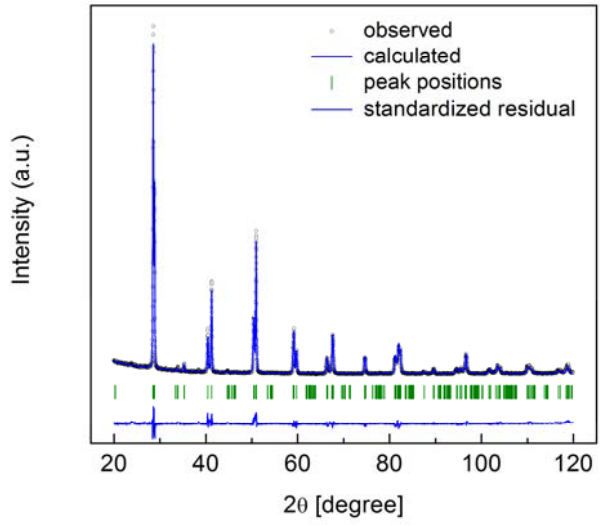

**Figure 2**

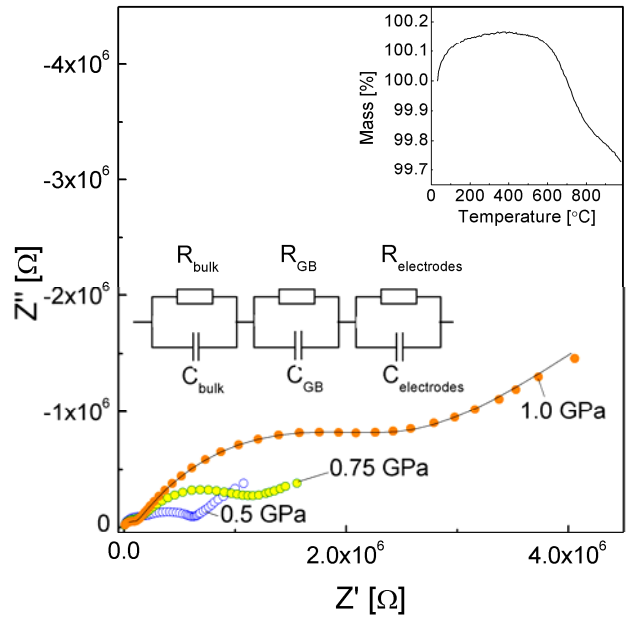

Figure 3

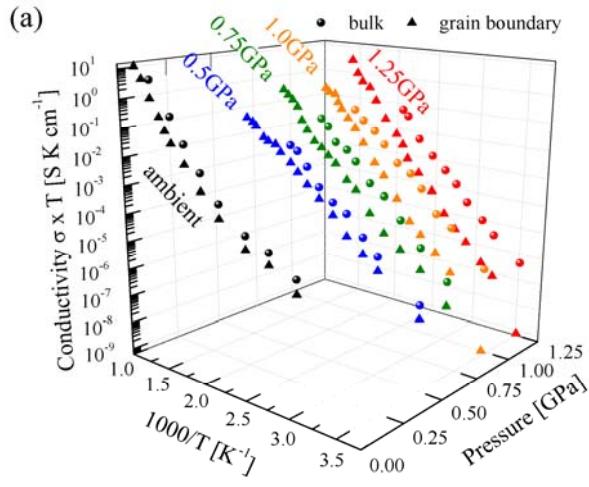

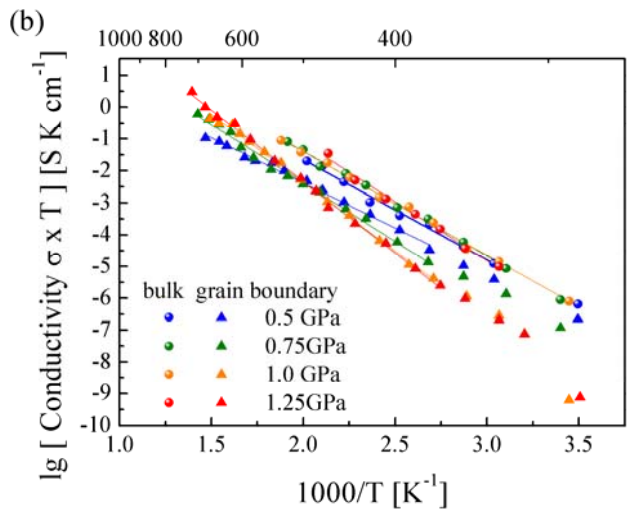

Figure 4

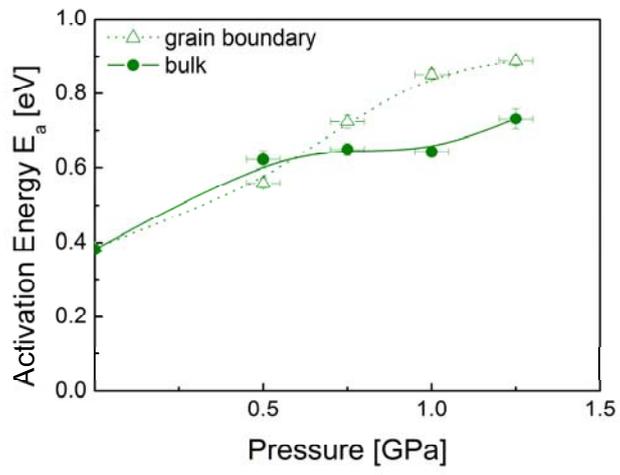

Figure 5

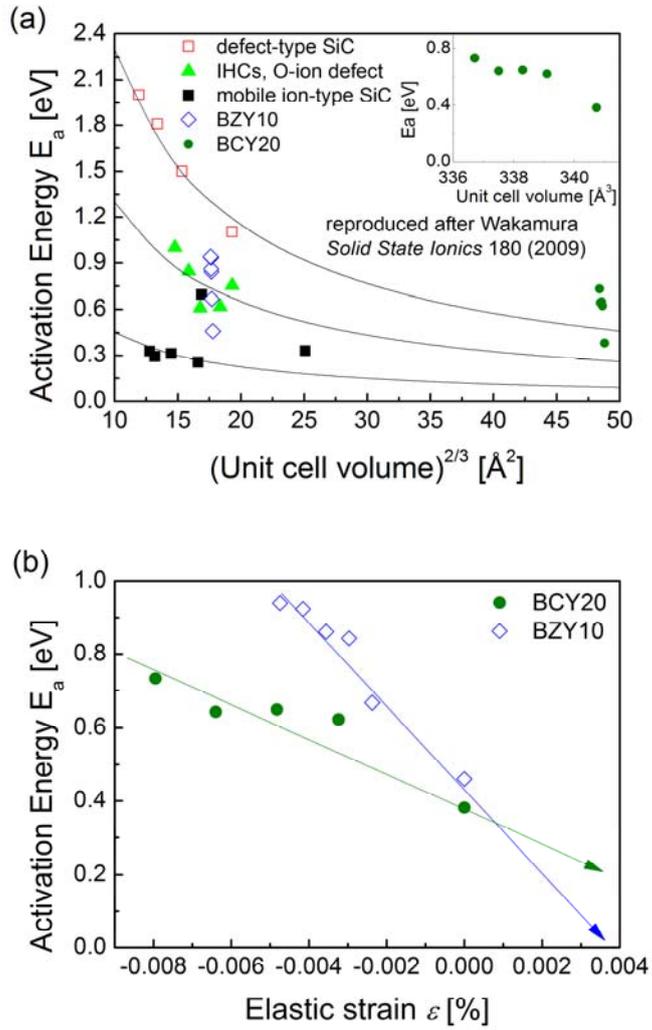